\documentclass[10pt]{amsart}
\usepackage{amsmath}
\usepackage{amssymb}
\usepackage{amsthm}
\usepackage{epsfig}
\usepackage{euscript}

\usepackage{amsfonts}

\newtheorem{thm}{Theorem}[section]

\newtheorem{dfn}[thm]{Definition}

\newtheorem{rk}[thm]{Remark}
\newcommand{\ndnt}{\hspace{7mm}}
\newcommand{\hs}{\hspace{2mm}}

\newcommand{\s}{\vspace{3mm}}

\begin{document}

\title[Toward Zeta Functions of Multifractals]
{Toward Zeta Functions and\\ Complex Dimensions of Multifractals}

\author[Lapidus and Rock]
{Michel~L.~Lapidus and John~A. Rock}
\date{\today}
\subjclass[2000]{Primary: 11M41, 28A12, 28A80. Secondary: 28A75,
28A78, 28C15} \keywords{Fractal string, geometric zeta function,
complex dimensions, multifractal measure, multifractal zeta
functions, partition zeta functions, Cantor set.}
\thanks{The work of the first author (M.L.L.) was partially supported by
the US National Science Foundation under the research grants
DMS-0707524 and (via the research program and workshop on ``Analysis
on Graphs and Fractals'' of the Isaac Newton Institute of
Mathematical Sciences in Cambridge, UK) under the grant
DMS-0648786.}

\begin{abstract}
Multifractals are inhomogeneous measures (or functions) which are
typically described by a full spectrum of real dimensions, as
opposed to a single real dimension. Results from the study of
fractal strings in the analysis of their geometry, spectra and
dynamics via certain zeta functions and their poles (the complex
dimensions) are used in this text as a springboard to define similar
tools fit for the study of multifractals such as the binomial
measure. The goal of this work is to shine light on new ideas and
perspectives rather than to summarize a coherent theory. Progress
has been made which connects these new perspectives to and expands
upon classical results, leading to a healthy variety of natural and
interesting questions for further investigation and elaboration.
\end{abstract}

\maketitle

\vspace{2cm}

\textsc{Michel L. Lapidus}\\
{\tiny \textsc{Department of Mathematics, University of
California,\\
Riverside, CA} 92521-0135 \textsc{USA} \par}

\textit{E-mail address:} \textbf{lapidus@math.ucr.edu}

\hspace{1cm}

\textsc{John A. Rock}\\
{\tiny \textsc{Department of Mathematics, California State
University, Stanislaus,\\ Turlock, CA} 95382 \textsc{USA} \par}

\textit{E-mail address:} \textbf{jrock@csustan.edu}

\pagebreak

\section{Introduction}\label{sum}

The concept of {\it multifractals} arises from the study of objects
in physics, geology, chemistry, economy and crystal growth (among
others) which are believed to be more fully described by a spectrum
of real dimensions rather than with a single real value such as the
Hausdorff or the Minkowski dimension. Chapter 9 of \cite{Sch}
provides an excellent introduction to some natural occurrences of
these multifractal objects, and Chapter 17 of \cite{Falc} and
Appendix B of \cite{PeitJS} provide mathematically  heuristic
descriptions of their construction and properties. Essentially, each
dimension in the spectrum corresponds in a specific manner to a
regularity value that describes the multi-scale behavior of the
object in question. In our case, we focus on a binomial measure on
the unit interval whose support is the ternary Cantor set.
Ultimately we would like to take into account the oscillations
intrinsic to such objects by allowing a spectrum of {\it complex
dimensions}, perhaps one for each regularity value.

\ndnt Section 2 features the construction of our primary example of
a multifractal (a binomial measure with support on the Cantor set)
and recalls some heuristic results on measures of this type from
\cite{Falc} and \cite{PeitJS}. Also, the definition of {\it
regularity} is recalled as it appears in \cite{LVT}. Regularity is
key to the description of the multi-scale behavior of multifractal
measures and functions.

\ndnt Section 3 provides a brief introduction to and a summary of
some results from the analysis of fractals in the study of the
geometry, spectra and dynamics of fractal strings via certain zeta
functions and their poles (the complex dimensions) in the classical
sense of \cite{LapPo1}, \cite{Lap2}, \cite{LapvF1}, and
\cite{LapvF4}. Further results and analysis from the theory of
complex dimensions of fractal strings can be found in
\cite{HL,LapPe1,LapPe2,LapvF1,LapvF4}.

\ndnt Section 4 summarizes some of the recent results from
\cite{LapRock} and \cite{Rock}. The definition of the {\it partition
zeta function} and some illustrative theorems on the connection to
current results from other approaches to multifractal analysis (see,
e.g., \cite{Ja1,KahPey,LapvF5,O1}) are given and discussed, in
particular the further solidification of the heuristic results
described in \cite{Falc} and \cite{PeitJS} and revisited in Section
2.

\ndnt Section 5 reviews the slightly less recent results (from
\cite{LLVR} and \cite{Rock}) on the use of multifractal zeta
functions to illuminate some topological properties of fractal
strings. These results generalize and expand the theory of complex
dimensions of ordinary fractal strings from \cite{LapvF1,LapvF4}.

\ndnt Section 6 closes with a collection of natural questions that
arise from the work and deserve further attention. Also, suggestions
for further research are given and briefly discussed.

\ndnt Overall, the goal of this work is to shine some light on new
ideas, not to summarize a coherent theory. The proofs and full
development of theorems have been left out for brevity but can be
found in the appropriate references.

\section{A Multifractal Measure on the Cantor Set}\label{nma}

\begin{figure}
\epsfysize=3.7cm\epsfbox{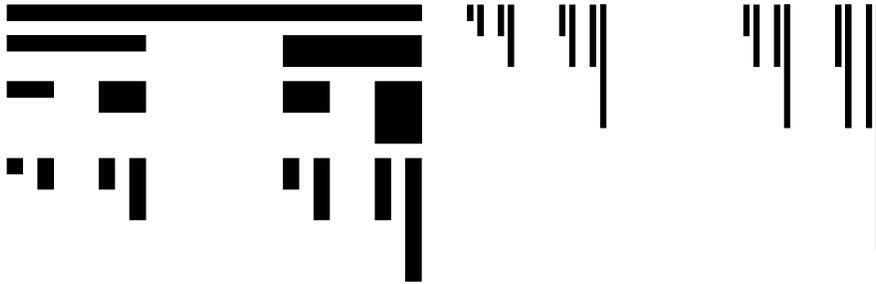}
    \caption{\textit{Constructing a binomial measure $\mu$ on the Cantor set.}}
\end{figure}

This section begins with the construction of a simple example of a
multifractal measure. A binomial measure $\mu$ can be constructed by
adding a mass distribution to the construction of the Cantor set
which consists of a countable intersection of a nonincreasing
sequence of closed intervals whose lengths tend to zero.
Specifically, in addition to removing open middle thirds, weight is
added at each stage. On the remaining closed intervals of each stage
of the construction, place $1/3$ of the weight on the left interval
and $2/3$ on the right, ad infinitum. See Figure 1. The measure
found in the limit, denoted $\mu$, is a multifractal measure.

\ndnt An integral notion of this text which stems from other
approaches to multifractal analysis is {\it regularity} (or {\it
coarse H\"older exponent}). This notion and some illustrative
theorems on its application to a mathematical formalism for
multifractals can be found in \cite{LVT}. In our case, as well as
that of \cite{LLVR,LapRock,LVM,LVT,Rock}, regularity allows for the
breakdown of a given multifractal measure by observing its behavior
at different scales.

\begin{dfn}\label{def:reg}
Let \( \textnormal{\textbf{X}([0,1])} \) denote the space of closed
subintervals of \([0,1]\). The \underline{regularity} \(A(U)\) of a
Borel measure $\mu$ with \( U \in \textnormal{\textbf{X}([0,1])} \)
and range $[0,\infty]$ is
\[
A(U)=\frac{\log \mu(U)}{\log{|U|}},
\]
where \( |\cdot| = \lambda(\cdot) \) is the Lebesgue measure on
$\mathbb{R}$.
\end{dfn}
Equivalently, $A(U)$ is the exponent $\alpha$ that satisfies
\[
|U|^{\alpha}=\mu(U).
\]
Note that regularity can be considered for any interval, whether
open, closed or neither.

\begin{figure}
\epsfysize=9cm\epsfbox{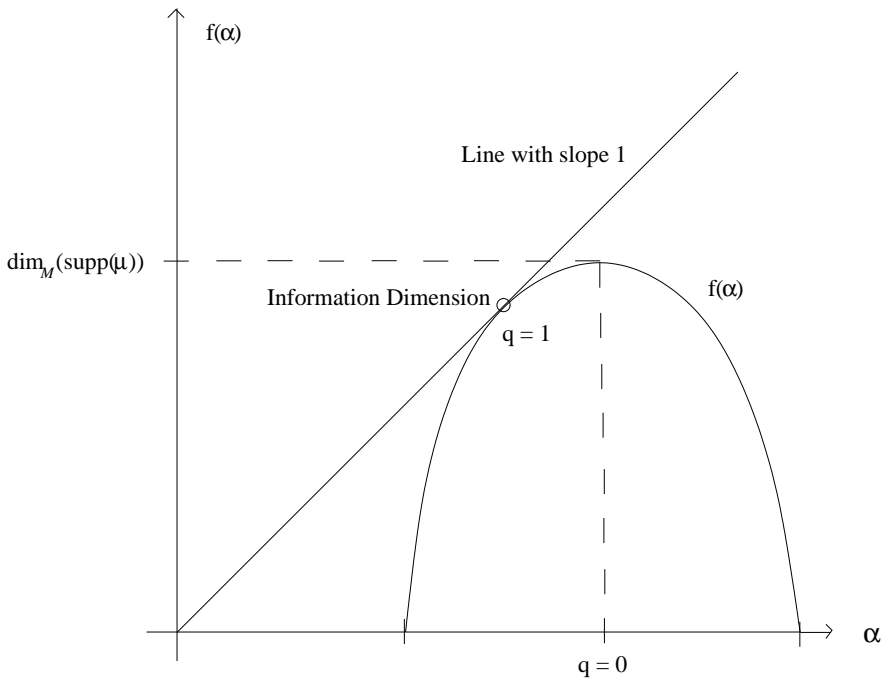}
    \caption{\textit{The multifractal spectrum curve of $f(\alpha)$ for the measure $\mu$, as found in \cite{Falc}, page 259.}}
\end{figure}

\ndnt Collecting intervals according to their regularity is key to
the developments of the multifractal and partition zeta functions
(defined in sections 5 and 4 respectively), which mirror that of the
geometric zeta functions (in section 3) in certain respects. Thus,
the following notation is helpful: Let \(U \in \mathcal{R}(\alpha)
\) if and only if \(A(U) = \alpha.\) Regularity values $\alpha$ in
the extended real numbers $[-\infty,\infty]$ will be considered. In
the extreme cases,
\[
\alpha=\infty=A(U) \Leftrightarrow \mu(U) =0 \textnormal{ and } |U|
> 0,
\]
and
\[
\alpha=-\infty=A(U) \Leftrightarrow \mu(U) = \infty  \textnormal{
and } |U| > 0.
\]

\ndnt Fixing the regularity $\alpha$ allows for the definition of a
generalization of geometric zeta functions called {\it multifractal
zeta functions}. These functions, originally from \cite{LLVR} and
discussed in Section \ref{mzf}, yield additional topological
information for fractal strings and suggest a way of conducting
multifractal analysis for Borel measures on the unit interval. The
properties of these multifractal zeta functions have not yet been
shown to relate significantly to current results in multifractal
analysis. However, as discussed in Section 4, the {\em partition
zeta functions} first described in \cite{Rock} and later in
\cite{LapRock} do relate to the results described in \cite{Falc} and
\cite{PeitJS} mentioned below. A similar approach to this type of
analysis appears in the work-in-progress \cite{LVM}, where very nice
connections to classical results also exist and are explained.

\ndnt Current preliminary results on the measure $\mu$ can be found
in Chapter 17 of \cite{Falc}. To briefly summarize, a Legendre
transform pair using the {\it multifractal spectrum} $f(\alpha)$ of
regularity $\alpha$ and parameter $q$ results in the curve in Figure
2. Roughly speaking, each regularity value $\alpha$ corresponds to a
collection of sets which have the same regularity value at
ever-decreasing scales and combine to yield a dimension (of sorts)
for that regularity value. In particular, when the parameter $q =1$,
the regularity value yields the information dimension of the measure
$\mu$, and when $q=0$, the regularity value yields the Minkowski
dimension of the support of $\mu$. This multifractal spectrum is
rediscovered as the abscissa of convergence function
$\sigma(\alpha)$ for the partition zeta functions of the measure
$\mu$, as described in Section 4 of this work.

\ndnt Multifractal analysis has not yet been defined in a common,
strict sense. Indeed, different authors provide a variety of
different approaches arising from both mathematics and applications.
For a physics perspective, see, e.g., \cite{Man,PF,Sch} and part of
\cite{LapvF5}. For a mathematical perspective, see Chapter 17 of
\cite{Falc} and Appendix B of \cite{PeitJS} for an introduction to
the subject matter, and see \cite{Sch} from an applications
standpoint. For even more on the mathematics side, see
\cite{Ja1,Ja2,Jaf,KahPey,LapvF5,BM,Ol,O1,O2}.

\ndnt Before elaborating on the connections between the results
described in this work and the results described in \cite{Falc} and
\cite{PeitJS}, the next section discusses the techniques used in the
study of fractal strings that motivate the definitions of the
multifractal and partition zeta functions.

\section{Fractal Strings}\label{fs}

A {\it fractal string}, in the classical sense of \cite{LapPo1},
\cite{Lap2}, \cite{LapvF1} and \cite{LapvF4}, is a bounded, open
subset of the real line denoted by \(\Omega =
\cup_{j=1}^{\infty}(a_j,b_j).\) Such objects consist of an at most
countable collection of disjoint open intervals. When there is a
countably infinite collection of disjoint intervals in $\Omega$
contained in $[0,1]$ and the boundary $\partial\Omega$ is equal to
the complement $\Omega^c$ in the usual topology of $[0,1]$, the
resulting set is often a fractal subset of the real line. An example
of such an object which is directly pertinent to this work is the
Cantor String. Its complement in $[0,1]$ is the classic Cantor set.
See Figure 3 for a finite approximation of the Cantor String.
Throughout this text\footnote{Theorem \ref{thm:recovergzf} is an
exception.}, we assume that a fractal string $\Omega$ has total
length 1, is a subset of $[0,1]$, and has boundary $\partial\Omega$
equal to $\Omega^c$, the complement of $\Omega$ in $[0,1]$.

\ndnt Important geometric, spectral and dynamical information is
contained in the collection of the lengths of the intervals which
constitute $\Omega$. This collection is denoted $\mathcal{L}$. Thus,
\( \mathcal{L} = \{\ell_j\}_{j=1}^{\infty}, \) where the $\ell_j$
are the lengths of the disjoint open intervals $(a_j,b_j)$. Often it
is best to consider $\mathcal{L}$ as a set of distinct lengths
$\{l_n\}_{n=1}^{\infty}$ with multiplicities
$\{m_n\}_{n=1}^{\infty}$. The results of this section depend only on
the sequence of lengths $\mathcal{L}$ and not on the topological
configuration of the disjoint open intervals of the fractal string
$\Omega$ which generate them. As such, and by abuse of notation, the
lengths $\mathcal{L}$ may be referred to as the fractal string in
place of the open set $\Omega$. A discussion regarding the
differences between fractal strings with varying topological
configuration but identical lengths can be found in Section 5. The
independence of the results presented in this section from the
topological configuration of the relevant fractal strings is a key
motivation for the definitions that follow in Section 4. The lengths
themselves retain much information regarding the fractal strings,
and a similar comment can be applied to the lengths associated with
the construction in Section 4 of a multifractal measure such as
$\mu$ from Section 2.

\ndnt In this work, the key notion of dimension is the {\it
Minkowski dimension}, whereas in many other works on fractals and
multifractals, the Hausdorff dimension is prominent. The following
develops the definition of the Minkowski dimension.

\ndnt The one-sided volume of the tubular neighborhood of radius
$\varepsilon$ of $\partial\Omega$ is
\begin{eqnarray}
V(\varepsilon)=\lambda(\{x \in \Omega \hs | \hs
dist(x,\partial\Omega)<\varepsilon\}),
\end{eqnarray}
where $\lambda(\cdot)=|\cdot|$ denotes the Lebesgue measure. The
\textit{Minkowski dimension} of $\partial\Omega$, or simply of
$\mathcal{L}$, is
\begin{eqnarray}
\dim_M(\partial\Omega)=D=D_{\mathcal{L}}:=\inf \{d \geq 0 \hs | \hs
\limsup_{\varepsilon \rightarrow
0^{+}}V(\varepsilon)\varepsilon^{d-1} <\infty \}.
\end{eqnarray}

Note that one may refer directly to the Minkowski dimension of the
sequence of lengths $\mathcal{L}$.

\begin{figure}
\epsfysize=1.2cm\epsfbox{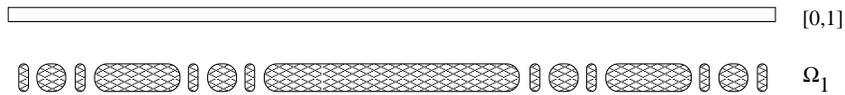}
    \caption{\textit{An approximation of the Cantor String $\Omega_1$.}}
\end{figure}

\ndnt The equation below describes a relationship between the
Minkowski dimension $D$ of (the boundary of) a fractal string with
lengths $\mathcal{L}$ and the sum of each of its lengths with
exponent $\sigma \in \mathbb{R}$. This was first observed in
\cite{Lap2} using a key result of Besicovitch and Taylor
\cite{BesTa}, and a direct proof can be found in \cite{LapvF4}, pp.
17--18.

\begin{eqnarray}
D=D_{\mathcal{L}}=\inf \left\{\sigma \in \mathbb{R} \hs | \hs
\sum_{j=1}^{\infty}\ell_{j}^{\sigma} <\infty \right\}.
\end{eqnarray}

$D_{\mathcal{L}}$ is the abscissa of convergence of the Dirichlet
series \(\sum_{j=1}^{\infty}\ell_{j}^{s}\), where $s \in
\mathbb{C}$. This Dirichlet series is the {\it geometric zeta
function} of $\mathcal{L}$ and it is the function that has been
generalized in \cite{LLVR,LapRock,Rock} using notions from
multifractal analysis.

\begin{dfn}\label{def:gzf} The \underline{geometric zeta function}
of a fractal string $\Omega$ with lengths $\mathcal{L}$ is
\[
\zeta_{\mathcal{L}}(s)=\sum_{j=1}^{\infty}\ell_{j}^{s}=
\sum_{n=1}^{\infty}m_{n}l_{n}^{s},
\]
where $\textnormal{Re}(s)>D_{\mathcal{L}}$.
\end{dfn}

When possible, such a function can be meromorphically continued to a
region $W$ in the complex plane. Such an extension may reveal a
family of poles which are called the {\it complex dimensions} of the
string $\mathcal{L}$, denoted $\mathcal{D}_{\mathcal{L}}$, which has
the following definition.

\begin{dfn}\label{def:cd}
The set of \underline{complex dimensions} of a fractal string
$\Omega$ with lengths $\mathcal{L}$ is
\[
\mathcal{D}_{\mathcal{L}}(W)=\{\omega \in W \hs | \hs
\zeta_{\mathcal{L}} \hs \textnormal{has a pole at } \omega\}.
\]
\end{dfn}

\ndnt Many interesting results stem from the investigation of the
geometric zeta functions and complex dimensions of fractal strings.
For instance, the following theorem characterizes the Minkowski
measurability of a fractal string and can be found in
\cite{LapvF1,LapvF4}. First, we define some of the terms used in the
theorem.

\ndnt If \(\lim_{\varepsilon \rightarrow
0^{+}}V(\varepsilon)\varepsilon^{d-1}\) exists and is positive and
finite for some $d$, then $d = D$ and we say that $\mathcal{L}$ is
\textit{Minkowski measurable}. The \textit{Minkowski content} of
$\mathcal{L}$ is then defined by
\(\mathcal{M}(D,\mathcal{L}):=\lim_{\varepsilon \rightarrow
0^{+}}V(\varepsilon)\varepsilon^{D-1}.\) The Minkowski
dimension\footnote{Defined as previously in Eq. (2), except 1 is
replaced with $r$.} can be expressed in terms of the upper box
dimension
\begin{eqnarray}
\limsup_{\varepsilon \rightarrow
0^{+}}\frac{N_{\varepsilon}(F)}{-\log{\varepsilon}},
\end{eqnarray}
where $N_{\varepsilon}(F)$ is the smallest number of cubes with side
length $\varepsilon$ that cover a nonempty bounded subset $F$ of
$\mathbb{R}^r$. In \cite{Lap1}, it is shown that if $F =
\partial\Omega$ is the boundary of a bounded open set $\Omega$, then
\(r-1 \leq \dim_H(F) \leq \dim_M(F) \leq r\) where $r$ is the
Euclidean dimension of the ambient space, $\dim_H(F)$ is the
Hausdorff dimension of $F$ and $\dim_M(F)=D$ is the Minkowski
dimension of $F$. In this work, we have $r=1$ and thus \(0 \leq
\dim_H(F) \leq \dim_M(F) \leq 1.\) The specific conditions under
which the theorem below holds are too complicated to concisely
describe in this work, however the full theorem and context can be
found in Chapter 8 of \cite{LapvF4}.

\begin{thm}\label{thm:cdmink}
If a fractal string $\Omega$ with lengths $\mathcal{L}$ satisfies
certain mild conditions, then the following statements are
equivalent:
\begin{enumerate}
\item $D$ is the only complex dimension of $\Omega$ with real part
$D_{\mathcal{L}}$, and it is simple.
\item $\partial \Omega$ is Minkowski measurable.
\end{enumerate}
\end{thm}

\ndnt The complex dimensions of the Cantor String are easily found.
The distinct lengths are $l_n = 3^{-n}$ with multiplicities $m_n =
2^{n-1}$ for every $n \in \mathbb{N}$. Hence, for
\(\textnormal{Re}(s) > \log_3{2},\)
\[
\zeta_{\mathcal{L}}(s)= \zeta_{CS}(s) =
\sum_{n=1}^{\infty}2^{n-1}3^{-ns} = \frac{3^{-s}}{1-2 \cdot 3^{-s}}.
\]

The last equation holds for all $s \in \mathbb{C}$ after analytic
continuation, hence
\[
\mathcal{D}_{\mathcal{L}} = \mathcal{D}_{CS} = \left\{ \log_3{2} +
\frac{2im\pi}{\log3} \hs | \hs m \in \mathbb{Z} \right\}.
\]

\begin{rk} \textnormal{Theorem \ref{thm:cdmink} applies to self-similar strings
(fractal strings whose boundary is a self-similar set). The Cantor
String is self-similar, thus Theorem \ref{thm:cdmink} indicates that
the Cantor String is {\it not} Minkowski measurable.}
\end{rk}

\ndnt The texts \cite{LapvF1,LapvF4} (specifically Chapter 8 of
\cite{LapvF4}) also contain the following key result, which uses the
complex dimensions of a fractal string in a formula for the volume
of the inner $\varepsilon$-neighborhoods of the fractal string.

\begin{thm}[Tube Formula]\label{thm:vol}
The volume of the one-sided tubular neighborhood of radius
$\varepsilon$ of the boundary of a fractal string $\Omega$ with
lengths $\mathcal{L}$ is given (under mild hypotheses) by the
following explicit formula with error term:

\[
V(\varepsilon) = \sum_{\omega \in \mathcal{D}_{\mathcal{L}}(W)\cup
\{0\}} \textnormal{res}\left(\frac{\zeta_{\mathcal{L}}
(s)(2\varepsilon)^{1-s}}{s(1-s)};\omega \right) +
\mathcal{R}(\varepsilon),
\]
where the error term can be estimated by \( \mathcal{R}(\varepsilon)
= \mathcal{O}(\varepsilon^{1-\sigma}) \) as \( \varepsilon
\rightarrow 0^+\) with $\sigma$ being the upper bound of the graph
of a certain bounded, real-valued continuous function. When the
horizontal and vertical axes are exchanged, the graph of this
function is the left part of the boundary of the region $W$.
\end{thm}

\begin{rk} \textnormal{If $\mathcal{L}$ is a self-similar string,
the hypotheses of Theorem \ref{thm:vol} are always satisfied and its
conclusion holds with} \( \mathcal{R}(\varepsilon) \equiv 0.\)
\textnormal{This is the case for the Cantor String, for example.}
\end{rk}

\ndnt In order to illustrate Theorem \ref{thm:vol} in a very simple
situation, we give the concrete form of the tube formula for the
Cantor String (see \cite[Eq.(1.14),~p.15]{LapvF4}):
\[
V_{CS}(\varepsilon) = \frac{1}{2\log3} \sum_{n= -\infty}^{\infty}
\frac{(2\varepsilon)^{1-D-in\textbf{p}}}{(D+in\textbf{p})(1-D-in\textbf{p})}
-2\varepsilon,
\]
for all $0 <  \varepsilon \leq 1/2,$ where $D= \log_{3}2$ is the
Minkowski dimension of the Cantor String (and the Cantor set) as
above, and $\textbf{p}= 2\pi /\log3$ is its oscillatory period.

\ndnt Important geometric and spectral information regarding the
structure of a fractal string is contained in its sequence of
lengths $\mathcal{L}$. The real parts of the complex dimensions are
related to the amplitudes of the oscillations in the volume
$V(\varepsilon)$ of the tubular neighborhood of the string, while
the imaginary parts coincide with the frequencies. Other examples
elaborating this philosophy are provided in \cite{LapvF1,LapvF4} as
well as \cite{LapPe1,LapPe2}, where some higher--dimensional
counterparts to Theorem \ref{thm:vol} in the case of the Koch
snowflake curve and self-similar systems (and tilings) can be found.

\ndnt The strength of these results and their independence from the
topological configuration of the fractal string $\Omega$, and hence
their complete dependence on the lengths $\mathcal{L}$ alone, is a
key motivation for defining the partition zeta functions in the next
section in terms of sequences of lengths with certain properties.
The next section translates the construction of the geometric zeta
function of a fractal string into the construction of a family of
partition zeta functions for a multifractal measure such as the
measure $\mu$ from Section 2.

\section{Partition Zeta Functions}\label{pzf}

The {\it partition zeta functions} described in this section were
first defined in \cite{Rock} and will be discussed further in
\cite{LapRock}. These functions are defined for any Borel measure on
the unit interval along with a family of partitions and are
parameterized by the regularity values attained by such measures.
The construction is very similar to that of the geometric zeta
function in the sense that the functions are series whose terms come
from a sequence of properly defined lengths. For geometric zeta
functions the terms are derived from the lengths of the disjoint
intervals of a given fractal string. However, for partition zeta
functions the terms are derived from a sequence of lengths from a
family of partitions which exhibit the same regularity.

\ndnt The families of partitions we consider satisfy certain
requirements and occur quite naturally in the construction of
multinomial measures such as the binomial measure. Consider an
ordered family of partitions $\mathfrak{P} =
\{\mathcal{P}_n\}_{n=1}^{\infty}$ of $[0,1]$, each of which splits
the unit interval into finitely many subintervals. The order is
given by the relation $\mathcal{P}_n \succ \mathcal{P}_{n+1}$ taken
to mean that each of the intervals $P_{n+1}^k$ which comprise the
partition $\mathcal{P}_{n+1}$ is a subinterval of some interval in
$\mathcal{P}_n$\footnote{To avoid trivial situations, we further
assume that the mesh of the sequence of partitions $\mathcal{P}_n$
tends to zero.}.

\begin{dfn}\label{def:pzf}
For a measure $\mu$ on the interval $[0,1]$ with an ordered family
of partitions $\mathfrak{P}$, the \underline{partition zeta function
with regularity $\alpha$} is
\[
\zeta^{\mu}_{\mathfrak{P}}(\alpha,s) =
\sum_{n=1}^{\infty}\sum_{A(P^k_n)=\alpha}|P^{k}_{n}|^s,
\]
where the inner sum is taken over the intervals $P^k_n$ in the
partition $\mathcal{P}_n$ which have regularity $A(P^k_n)=\alpha \in
[-\infty,\infty]$, and $\textnormal{Re}(s)$ is large enough.
\end{dfn}

\ndnt Recall the construction of the measure $\mu$ from Section 2.
The breakdown of mass and length readily defines a {\it natural
family of partitions} $\mathfrak{P}$ for this measure $\mu$ as
simply the closed intervals and their complements in the
construction of the Cantor set. The mass breakdown allows for the
separation of the individual intervals in $\mathfrak{P}$ into
collections according to their regularity, found with respect to the
measure $\mu$. At each stage, the intervals with the same regularity
$\alpha$ have multiplicities given by binomial coefficients. In
turn, their lengths and multiplicities constitute the terms in the
definition of the partition zeta function with that regularity.

\ndnt To determine the intervals which have the same regularity,
note that the collection of all intervals from every partition in
the family $\mathfrak{P}$ is a countable set. Thus, the regularity
values attained on these intervals are a function of the ordered
pair of integers $(k_1,k_2)$ (which satisfy the properties mentioned
below) as follows:
\[
\alpha := \alpha(k_1, k_2) =
\frac{\log{(2^{nk_1}/3^{nk_2})}}{\log{(1/3^{nk_2})}} = 1 -
\frac{k_1}{k_2}\log_3{2},
\]
for all $n \in \mathbb{N}$ and where $k_1 \in \mathbb{N} \cup
\{0\}$, $k_2 \in \mathbb{N}$, $k_1 \leq k_2$ and $k_1$ and $k_2$ are
necessarily relatively prime (denoted $(k_1,k_2)=1$), except when
$k_1 = 0$ or $1$ and $k_2 = 1$. The integers $k_j$ relate to the
measure and regularity of intervals in that, roughly, $k_1$ is the
number of times an interval falls to the right and gets $2/3$ of the
mass after $k_2$ stages during the construction.

\ndnt The regularity value $\alpha(k_1,k_2)$ with fixed $(k_1,k_2)$
as above only occurs in the partitions $\mathcal{P}_{nk_2}$ for each
$n \in \mathbb{N}$, with multiplicity $\binom{nk_2}{nk_1}$. In
summation,
\[
\zeta^{\mu}_{\mathfrak{P}}(\alpha(k_1,k_2),s) =
\zeta^{\mu}_{\mathfrak{P}}(\alpha(k_2-k_1,k_2),s) =
\sum_{n=1}^{\infty} \binom{nk_2}{nk_1} 3^{-k_2ns}.
\]
See Figure 4 for the first several intervals with regularity
$\alpha{(1,2)}$.

\begin{figure}
\epsfysize=3.7cm\epsfbox{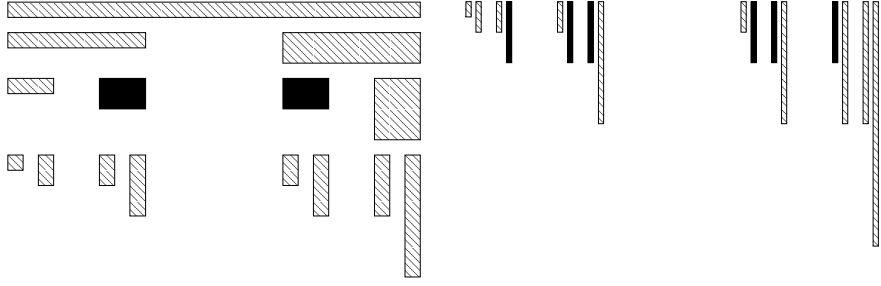}
    \caption{\textit{Construction of the multifractal binomial
    measure $\mu$, with emphasis on the intervals with regularity
    $\alpha(1,2)$.}}
\end{figure}

\ndnt There is a notion of multifractal spectrum which stems
immediately from this set up and is reminiscent of similar results
on multifractals found in Chapter 17 of \cite{Falc}, especially the
graphs of the spectra in Figures 17.2 on page 259 (reproduced in
Section 2 of this work) and 17.3 on page 261. In our context, the
spectrum $\sigma(\alpha)$ is defined as the function which yields
the abscissa of convergence of the partition zeta function with
regularity $\alpha$, $\zeta^{\mu}_{\mathfrak{P}}(\alpha,s)$, for all
$\alpha$. See Figure 5 for an approximation of the graph for this
function and compare to Figure 2 which contains the graph of the
multifractal spectrum $f(\alpha)$ from \cite{Falc}.

\ndnt This graph exhibits some interesting properties, such as its
maximum coincides with the Minkowski dimension of the support of
$\mu$. Further, this structure holds in greater generality. If, in
the construction of the Cantor set, the initial length is replaced
by some $h^{-1}$ and the smaller weight by a $w^{-1}$, we have the
following theorem. The proof is omitted, but can be found in
\cite{LapRock,Rock}.

\begin{thm}\label{thm:max} As a function of the regularity values,
the abscissa of convergence function $\sigma$ associated with the
partition zeta function of the measure $\mu(h,w)$ with $h\geq2$ and
$w>2$ has the form
\begin{eqnarray}
\sigma(\alpha) &=& \frac{(\alpha - \log_h{w})}{\log_h{(w-1)}} \cdot
\log_h\left( \frac{-(\alpha - \log_h{w})}{\log_h{(w-1)} }\right)\\
& &- \left(1+ \frac{(\alpha - \log_h{w})}{\log_h{(w-1)}}\right)
\cdot \log_h{\left(1+ \frac{(\alpha - \log_h{w}) }{
\log_h{(w-1)}}\right) }.\notag
\end{eqnarray}
\ndnt  As the abscissa of convergence function, $\sigma$ is defined
on a dense subset of the interval $[\log_h{w}-\log_h{(w-1)},
\log_h{w}]$, and it attains its maximum at
\begin{eqnarray}
\alpha =
\alpha(1,2) = \log_h{w}-(1/2)\log_h{(w-1)}.
\end{eqnarray}
This maximum value coincides with the Minkowski dimension of the
support of the measure $\mu(h,w)$. That is,
\begin{eqnarray}
\dim_M(supp(\mu(h,w))&=& \max\{\sigma(\alpha) \hs | \hs \alpha =
\alpha(k_1,k_2), (k_1,k_2)=1\}\\
&=& \log_h{2}.\notag
\end{eqnarray}
\end{thm}

\begin{figure}
\epsfysize=10cm\epsfbox{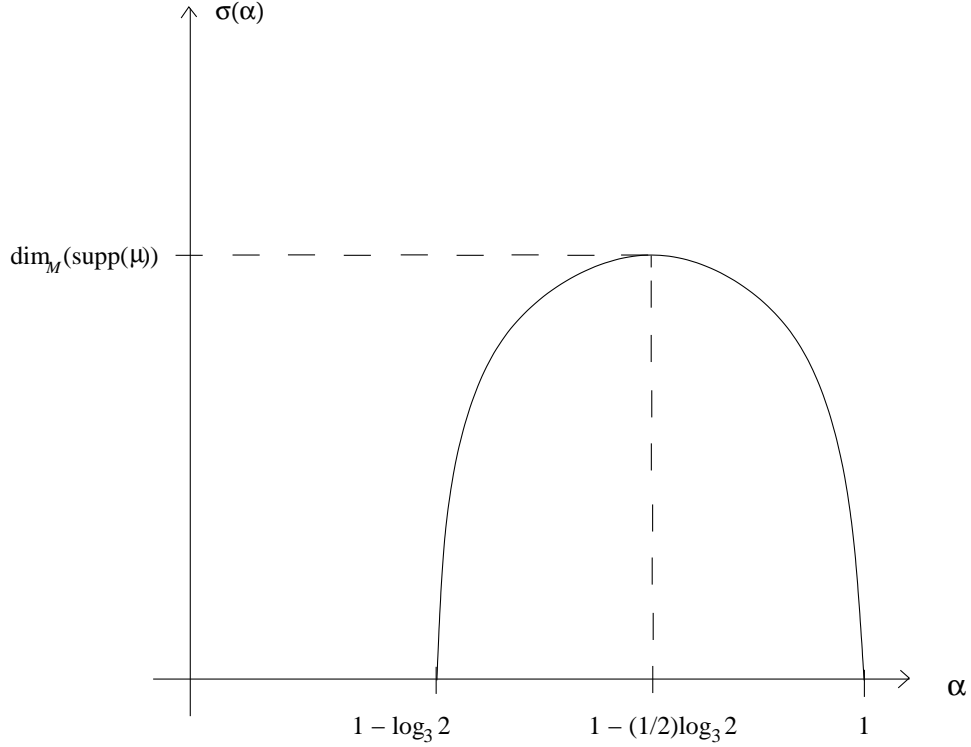}
    \caption{\textit{$\sigma$ as a function of $\alpha$ for the measure $\mu$.}}
\end{figure}

\ndnt Theorem \ref{thm:max} almost contains the following
multifractal spectrum $f(\alpha)$ discussed on page 934 of Appendix
B in \cite{PeitJS} as a specific case:

\begin{eqnarray*}
\sigma(\alpha) = f(\alpha) =
& &-\frac{\alpha_{max}-\alpha}{\alpha_{max}-\alpha_{min}}\log_2\left(\frac{\alpha_{max}-\alpha}{\alpha_{max}-\alpha_{min}}\right)\\
            & &-\frac{\alpha-\alpha_{min}}{\alpha_{max}-\alpha_{min}}\log_2\left(\frac{\alpha-\alpha_{min}}{\alpha_{max}-\alpha_{min}}\right),
\end{eqnarray*}
where $\alpha_{max}$ and $\alpha_{min}$ denote the maximum and
minimum regularity values attained by the measure in question. To
fit the measure from page 934 of Appendix B in \cite{PeitJS} to our
setting, take $h=2$ and $w=3$. The pertinent regularity values
become
\[
\alpha := \alpha(k_1, k_2) =
\frac{\log{(2^{nk_1}/3^{nk_2})}}{\log{(1/2^{nk_2})}} = \log_2{3} -
\frac{k_1}{k_2},
\]
where $\alpha_{max}$ and $\alpha_{min}$ are attained when $k_1 = 0$
with $k_2=1$ and $k_1 =k_2 =1$, respectively. We say that $f$ is
``almost'' a specific case because the equation $\sigma(\alpha) =
f(\alpha)$ only holds on a dense and discrete subset of
$[\alpha_{min},\alpha_{max}]$, as opposed to the full interval on
which $f$ is defined.

\ndnt In addition, the graph of the spectrum $f(\alpha)$ depicted in
Section 2 of this work and Chapter 17 of \cite{Falc} also stems from
an equation such as (5) from Theorem \ref{thm:max} (see page 261 of
\cite{Falc}).

\ndnt Among the common features of these graphs are the coincidence
of the maximum height of the curve with the Minkowksi dimension of
the support of the underlying measure and the symmetry about the
vertical line that passes through this maximum height. The symmetry
of $\sigma(\alpha)$ is evident from the equality of the binomial
coefficients $\binom{nk_2}{nk_1}$ and $\binom{nk_2}{n(k_2-k_1)}$ in
the respective partition zeta functions.

\ndnt The distinctions between the abscissa of convergence function
$\sigma$ of this text and the multifractal spectrum $f$ of
\cite{Falc} and \cite{PeitJS} lie in their developments. The
function $\sigma$ follows directly from the partition zeta functions
defined by the weighted partitions which define the measure $\mu$,
whereas $f$ follows from the same heuristic development but takes
its values from an appropriate Legendre transform.

\ndnt Further generalizations to measures with multiplicative
structure similar to that of $\mu$ have been made, but for brevity
we shall merely mention their existence. Also, in \cite{LVM},
another type of zeta function which describes multifractal measures
in a manner very similar to the partition zeta functions has been
defined and investigated. The results contained in that paper
provide even more connections between the analysis of fractal
strings via zeta functions and this new approach to multifractal
analysis.

\ndnt It is important to note that the partition zeta functions do
not yield the geometric zeta function as some kind of special case.
Indeed, there is no underlying fractal string or closed set that
results from a construction like that of the Cantor set for the
intervals of $\mathfrak{P}$ and a fixed regularity. So, although
they do not recover the geometric zeta function for fractal strings,
the partition zeta functions provide some interesting information
for multifractal measures, further solidifying the existing results
described, for example, in \cite{Falc} and \cite{PeitJS}.

\ndnt The next section describes a precursor to the partition zeta
function and the similar zeta function in \cite{LVM}. Despite its
name, the multifractal zeta function does not connect to
multifractal analysis as thoroughly as these other zeta functions,
but it is a generalization of the geometric zeta functions of
fractal strings and provides topological information which can not
be obtained from their complex dimensions.

\section{Multifractal (or Topological) Zeta Functions}\label{mzf}

The multifractal zeta function, which made its first appearance in
\cite{LLVR}, was initially developed to investigate the properties
of multifractal measures. Its definition also relies on the notion
of regularity, but the lengths come from a much larger and more
complicated family than the family given by $\mathfrak{P}$ for the
partition zeta functions. This larger collection creates many
computational and theoretical difficulties, yet two regularity
values ($\pm \infty$) yield new and existing results for fractal
strings. The results presented in this section are originally from
\cite{LLVR} (viewed from a slightly different perspective), and can
also be found in \cite{Rock} (viewed from the current perspective of
this work).

\ndnt In order to bring fractal strings into a framework that uses
regularity, an appropriate measure must be defined. For a fractal
string $\Omega$ in the unit interval, the measure $\mu_{\Omega}$ is
the measure which has a unit point-mass at every endpoint of the
fractal string. Thus, any interval which does not contain an
endpoint of $\Omega$ does not have mass, hence its regularity is
$\infty$. On the other hand, any interval which contains a
neighborhood of a limit point of the endpoints of $\Omega$ has
infinite mass, hence its regularity is $-\infty$. These measures
combine with the multifractal zeta functions to recover and extend
the results obtained for fractal strings through the geometric zeta
functions.

\ndnt Using intervals whose lengths appear in a sequence
$\mathcal{N}$ (which decreases to zero) and collecting them
according to their regularity $\alpha$ allow for the definition of
multifractal zeta function given below, where $k_{n}(\alpha)$ is the
number of new disjoint intervals $K^{n}_{p}(\alpha)$ which arise at
stage $n$. The intervals $K^{n}_{p}(\alpha)$ do not necessarily have
length in $\mathcal{N}$, rather they are the disjoint intervals of
the set which is the union of all closed intervals of length $\eta_n
\in \mathcal{N}$ and the same regularity $\alpha$.

\begin{dfn}\label{def:mzf} The \underline{multifractal zeta function} of a measure
$\mu$, sequence $\mathcal{N}$, associated regularity value $\alpha
\in [-\infty,\infty]$ and is given by
\[
\zeta^{\mu}_{\mathcal{N}}(\alpha,s) =
\sum_{n=1}^{\infty}\sum_{p=1}^{k_{n}(\alpha)}|K^{n}_{p}(\alpha)|^s
\]
for \textnormal{Re}$(s)$ large enough.
\end{dfn}

\begin{figure}
\epsfysize=3cm\epsfbox{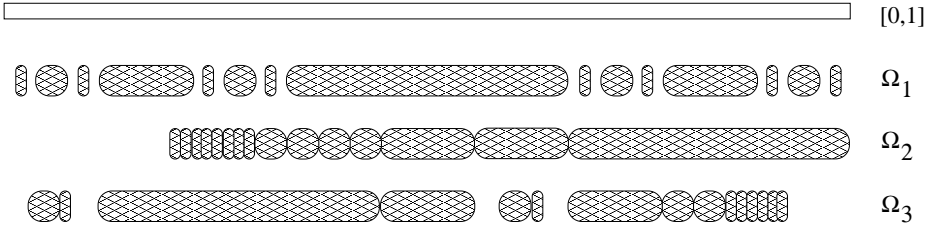}
    \caption{\textit{Three strings with the same lengths
    $\mathcal{L}$, the lengths of the Cantor String $\Omega_1$,
    but different topological configuration. }}
\end{figure}

\ndnt In this setting, we have the following theorem. The full proof
can be found in \cite{LLVR,Rock}. The basic idea of the proof is
that an interval with regularity $\alpha = \infty$ has no mass, thus
this interval must be a subset of the complement of the support of
the measure, the fractal string $\Omega_{\mu} =(supp(\mu))^c$. The
decreasing sequence $\mathcal{N}$ ensures that every disjoint open
interval in this fractal string is recovered, in turn enabling us to
recover the geometric zeta function. Unlike some of the other
results on fractal strings mentioned in this work, the following
theorem does not require the fractal string to have total length 1,
to be a subset of $[0,1]$, nor to have boundary equal to the
complement in the smallest compact interval which contains the
fractal string.

\begin{thm}\label{thm:recovergzf}
The multifractal zeta function of a positive Borel measure $\mu$,
any sequence $\mathcal{N}$ such that \(\eta_{n} \searrow 0\) and
regularity $\alpha = \infty$ is the geometric zeta function of
$\Omega_{\mu} =(supp(\mu))^c$ (where the complement is taken in the
smallest compact interval containing $supp(\mu)$), with lengths
$\mathcal{L}_{\mu}$. That is, \( \zeta^{\mu}_{\mathcal{N}}(\infty,s)
= \zeta_{\mathcal{L}_{\mu}}(s). \)
\end{thm}

\ndnt When a fractal string which has a boundary (complement in the
unit interval) that is a perfect set, such as the Cantor String, we
get the following theorem with omitted proof. The full development
and proof can be found in \cite{LLVR,Rock}. The regularity value
$-\infty$ allows us to distinguish between fractal strings with
identical lengths $\mathcal{L}$ and, hence, the same Minkowski
dimension, but with different topological arrangements. See Figure 6
for approximations of three fractal strings which have the same
$\mathcal{L}$ (the lengths of the Cantor String $\Omega_1$), but
have obviously distinct topological properties. In light of the
following theorem, one may refer to the multifractal zeta function
of a measure $\mu_{\Omega}$ with regularity $\alpha = -\infty$ as
the {\it topological} zeta function of the fractal string $\Omega$.

\begin{thm}\label{thm:perfect}
Let $\Omega$ be a fractal string with sequence of lengths
$\mathcal{L}$ and perfect boundary. Suppose that $\mathcal{N}$ is a
sequence which decreases to zero such that \(l_n>\eta_n \geq
l_{n+1}\) and \(l_n>2\eta_n,\) for all \(n \in \mathbb{N}.\) Then

\begin{eqnarray}
\zeta^{\mu_{\Omega}}_{\mathcal{N}}(\infty,s) &=&
\zeta_{\mathcal{L}}(s) = \sum_{n=1}^{\infty}m_nl_{n}^{s}, \textnormal{ and}\\
%\zeta^{\mu_{\Omega}}_{\mathcal{N}}(0,s) &=&
%\sum_{n=1}^{\infty}2m_n\eta_n^s, \textit{ and }\\
\zeta^{\mu_{\Omega}}_{\mathcal{N}}(-\infty,s) &=&
h(s)+\sum_{n=2}^{\infty}m_n(l_n-2\eta_{n})^s,
\end{eqnarray}

where $h(s)$ is the entire function given by \(h(s)=
\sum_{p=1}^{k_1(-\infty)}|K^{1}_{p}(-\infty)|^s. \)
\end{thm}

\ndnt For $q=1,2,3$ and fractal strings $\Omega_q$, Theorem
\ref{thm:perfect} yields

\[
\zeta^{\mu_q}_{\mathcal{N}}(\infty,s) = \zeta_{CS}(s) =
\frac{3^{-s}}{1-2\cdot3^{-s}}.
\]

However, the multifractal zeta functions corresponding to the
regularity value $-\infty$ have the following forms, where the
function $h_3$ is entire:

\begin{eqnarray*}
\zeta^{\mu_1}_{\mathcal{N}}(-\infty,s) &=&
    2\left( \frac{1}{3}+\frac{1}{9} \right)^s
    +\sum_{n=2}^{\infty} 2^{n-1} \left(
    \frac{1}{3^n}-\frac{2}{3^{n+1}} \right)^s\\
&=& 2\left( \frac{4}{9} \right)^s
    +\frac{2}{27^s} \left(\frac{1}{1-2\cdot3^{-s}} \right),\\
\zeta^{\mu_2}_{\mathcal{N}}(-\infty,s) &=& \eta_{1}^s =
\frac{1}{9^s},\\
 \zeta^{\mu_3}_{\mathcal{N}}(-\infty,s) &=& h_3(s)
    +\sum_{n=2}^{\infty}m_n \left(l_{2n-1}+l_{2n}
    -2\eta_{2n-1}\right)^s\\
&=& h_3(s) +\sum_{n=2}^{\infty}2^{n-1} \left( \frac{1}{3^{2n-1}}
    + \frac{1}{3^{2n}} - \frac{2}{3^{2n}}\right)^s\\
&=& h_3(s)+ \left(\frac{2^{s+1}}{81^s}\right) \left(
    \frac{1}{1-2\cdot9^{-s}} \right).\\
\end{eqnarray*}

\ndnt More definitively, it follows from the above discussion that
the poles (complex dimensions) of these multifractal zeta functions
differ completely:

\begin{eqnarray*}
\mathcal{D}^{\mu_1}_{\mathcal{N}}(-\infty) &=& \mathcal{D}_{CS} =
\left\{ \log_3{2} +
\frac{2i\pi m}{\log3} \hs | \hs m \in \mathbb{Z} \right\},\\
\mathcal{D}^{\mu_2}_{\mathcal{N}}(-\infty) &=& \emptyset,
\textnormal{\ndnt and}\\
\mathcal{D}^{\mu_3}_{\mathcal{N}}(-\infty) &=& \left\{ \log_9{2}+
\frac{2i\pi m}{\log9} \hs | \hs m \in \mathbb{Z} \right\}.
\end{eqnarray*}

\ndnt Thus, multifractal zeta functions for at least two regularity
values provide useful information about the properties of fractal
strings when certain measures are considered, specifically the
measures which have unit mass at every endpoint of the disjoint open
intervals which define the fractal string. These multifractal zeta
functions were the starting point for the development of the more
refined and relevant (with respect to multifractal analysis)
partition zeta functions.

\section{Conclusion}\label{conclusion}

There are many questions that arise in this new investigation of the
application of zeta functions to fractal and multifractal analysis.
For instance, when and where do the partition zeta functions have a
meromorphic extension? And what are their poles? Furthermore, what
are the ramifications? In light of the results with the complex
dimensions of fractal strings in \cite{LapvF1,LapvF4}, can we
capture other oscillations intrinsic to multifractals in terms of
these zeta functions? Can the theory be extended to
higher-dimensional multifractals, as was done in the case of
ordinary self-similar fractals in \cite{LapPe1,LapPe2}? And what
about multifractals which are not the result of a multiplicative
process?

\ndnt Although the multifractal zeta functions do not (conveniently)
provide significant information regarding a multifractal analysis of
measures, they allow for new results on the {\it topological}
properties of fractal strings to be obtained. In particular, such
results can not be obtained through use of the geometric zeta
functions of ordinary fractal strings alone. With this in mind, can
we randomize the theory in order to deal with more realistic
examples from the point of view of applications, as was done in
\cite{HL} for ordinary fractal strings?

\ndnt In general, what other information can be uncovered in the
world of fractals by means of such zeta functions?

\vspace{5mm}

{\it Acknowledgements.} The first author (M.L.L.) would like to
acknowledge the hospitality of the Institut des Hautes Etudes
Scientifiques (IHES), where he was a visiting professor while this
paper was completed.

\hspace{1cm}

{\tiny
\textsc{Michel L. Lapidus, }\textbf{lapidus@math.ucr.edu}\\
\textsc{Department of Mathematics, University of California,\\
Riverside, CA} 92521-0135 \textsc{USA} \par}

\s

{\tiny \textsc{John A. Rock, }\textbf{rock@csustan.edu}\\
\textsc{Department of Mathematics, California State University,
Stanislaus,\\ Turlock, CA} 95382 \textsc{USA} \par}

\end{document}